\documentclass[journal]{IEEEtran}

\ifCLASSINFOpdf
\else
   \usepackage[dvips]{graphicx}
\fi
\usepackage{url}

\usepackage{cite}
\usepackage{graphicx}
\usepackage{amsmath}
\usepackage{amssymb}
\usepackage{array}
\usepackage{float}
\usepackage{enumitem}
\usepackage{xcolor}
\usepackage{amsfonts}
\usepackage[ruled,linesnumbered]{algorithm2e}

\begin{document}

\title{One-Bit Compressive Sensing:\\ Can We Go Deep and Blind?}

\author{Yiming Zeng, \IEEEmembership{Student Member, IEEE}, Shahin Khobahi, \IEEEmembership{Student Member, IEEE},\\ Mojtaba Soltanalian, \IEEEmembership{Senior Member, IEEE} \thanks{This work was supported in part by the National Science Foundation Grant CCF-1704401.}
\thanks{The authors are with the Department of Electrical and Computer Engineering, University of Illinois at Chicago, Chicago, IL 60607, USA. \emph{Corresponding author: Yiming Zeng. E-mail: \emph{yzeng25@uic.edu}.}}
}


\markboth{IEEE SIGNAL PROCESSING LETTERS, 2022}
{Shell \MakeLowercase{\textit{et al.}}: Bare Demo of IEEEtran.cls for IEEE Journals}
\maketitle

\begin{abstract}
 One-bit compressive sensing is concerned with the accurate recovery of an underlying sparse signal of interest from its one-bit noisy measurements. The conventional signal recovery approaches for this problem are mainly developed based on the assumption that an exact knowledge of the sensing matrix is available. In this work, however, we present a novel data-driven and model-based methodology that achieves \emph{blind} recovery; i.e., signal recovery without requiring the knowledge of the sensing matrix. To this end, we make use of the deep unfolding technique and develop a model-driven deep neural architecture which is designed for this specific task. The proposed deep architecture is able to learn an alternative sensing matrix by taking advantage of the underlying unfolded algorithm such that the resulting \emph{learned} recovery algorithm can accurately and quickly (in terms of the number of iterations) recover the underlying compressed signal of interest from its one-bit noisy measurements. In addition, due to the incorporation of the domain knowledge and the mathematical model of the system into the proposed deep architecture, the resulting network benefits from enhanced interpretability, has a very small number of trainable parameters, and requires very small number of training samples, as compared to the commonly used black-box deep neural network alternatives for the problem at hand.

\end{abstract}

\vspace{6pt}


\IEEEpeerreviewmaketitle

\section{Introduction}

\IEEEPARstart{C}{ompressive} sensing (CS) is a  sampling framework that utilizes the frequently-encountered sparse nature of the underlying signals to overcome the limitations of the Nyquist and other traditional sampling paradigms \cite{4472240}. Within this framework, far fewer  samples are needed to maintain a high accuracy in signal reconstruction---in effect, leading to significant improvements in   sensing and signal recovery performance in various sparse settings. It is thus no surprise that CS has attracted a great deal of interest from many researchers due to its wide range of applications; see, e.g., \cite{5437428, 6674179, 4524050, gao2018compressive}, and the references therein.
\vspace{7pt}

In practice, the sampled signal needs to be quantized to certain levels for convenient data storage, transmission, and processing. One-bit CS \cite{4558487,7434599,plan2013one,8025559,7447738,mazumdar2021support} is a CS formulation that takes such a quantization to its extreme, where the quantizer is essentially a comparator that encodes the signal into one-bit of information $\{+1,-1\}$ based on its instantaneous level which is compared to a given threshold level. Employing such a comparator can significantly reduce the complexity of the sampling hardware  and achieves a faster data-acquisition speed \cite{9557819}. However, using the one-bit quantization model inevitably introduces additional noise to the acquired signal, which renders the accurate recovery of the underlying signal  a challenging task. 

The existing CS reconstruction techniques can be classified into two main categories of data-driven techniques that require sufficient training data for neural networks and model-based techniques with good interpretability. For instance, in \cite{yang2018admm,lu2018convcsnet,iliadis2018deep}, the authors consider the CS reconstruction problem from a deep learning viewpoint and develop data-driven techniques for reconstruction purposes. Particularly,  the authors of \cite{lu2018convcsnet} proposed a data-driven deep neural network based on two-branch convolutional neural networks (CNNs) to reconstruct image signals. Moreover, in \cite{wu2019deep}, the authors presented a framework for combining CS and deep networks via meta learning to jointly train the measurement and generation functions. 
On the other hand, several effective model-based CS algorithms have garnered popularity, including the \textit{Renormalized Fixed-Point Iteration} (RFPI) method \cite{movahed2012robust}, \textit{Matching Pursuit} (MP) \cite{tropp2007signal}, \textit{Binary Iterative Hard Thresholding} (BIHT) \cite{jacques2013robust}, and their extended versions. 
A main drawback of these model-based algorithms, however, is that a full knowledge of the sensing matrix is usually required during the recovery process. In the absence of such knowledge, the sensing matrix must first be recovered through a separate estimation module, upon which the estimated sensing matrix will be employed for signal reconstruction purposes.  
Specifically, the problem of one-bit blind compressive sensing was first considered in the seminal work \cite{zayyani2015dictionary} in which the authors proposed a dictionary learning algorithm to learn the underlying unknown sparsity domain. A similar problem was considered in \cite{wormann2013analysis} in which the authors utilize the geometric conjugate gradient algorithm to learn the sensing matrix in an adaptive manner and during the signal reconstruction process.

Additionally, a prominent problem in CS is the presence of noise, causing difficulties that have led many existing works in the field to assume a noise-free scenario. 
This problem is in fact exacerbated in the realm of one-bit data-acquisition where the quantized values are much more prone to additive noise near the threshold values. 
From a practical perspective, blind CS algorithm can recover dynamic magnetic resonance images from undersampled measurements \cite{6488855}. Also, learning a task-specific sensing matrix adaptively allows for an accurate recovery of sparse signal in the presence of noise.

\nocite{zayyani2015dictionary}

In this paper, we propose a hybrid model-based and data-driven method for the problem of blind one-bit compressive sensing based on the deep unfolding methodology \cite{hershey2014deep}, which can both learn the task-specific sensing matrix and recover the sparse signals of interest. This problem has been known for its difficulty even when high-resolution samples are available --- let alone the highly-quantized one-bit setting \cite{6034717}. 

\section{Modeling and Problem Formulation}
\subsection{One-Bit Compressive Sensing Model}

The one-bit compressive sensing data-acquisition model in a noise-free scenario can be formulated as follows:
\begin{equation}
    \boldsymbol{y}= f_\Theta(\boldsymbol{x})=\operatorname{sign}(\boldsymbol{\Phi} \boldsymbol{x}-\boldsymbol{\boldsymbol{\tau}}), \label{XX}
\end{equation}
where $\operatorname{sign}(x)= 1$ if $x \geq 0$, and $\operatorname{sign}(x)= -1$ otherwise, $\boldsymbol{\Phi}^{m\times n}$ represents the underlying sensing matrix, $\boldsymbol{x} \in \mathbb{R}^n$ is a $K$-sparse signal and $\boldsymbol{\tau}$ denotes the quantization threshold vector which pave the way for amplitude recovery. Here and in the rest of this work, we assume that all the functions are applied element-wise on the vector arguments.
We refer the interested reader to consult \cite{4558487,7434599}, and the references therein, for more detailed explanations on the one-bit CS paradigm and related models.

Our main goal in one-bit CS is to recover the $K$-sparse signal of interest $\boldsymbol{x} \in \mathbb{R}^n$ from the one-bit measurements $\boldsymbol{y}\in \mathbb{R}^m$ at the decoder. Based on the availability of information on the sensing variables, the below settings can be considered. 
\begin{itemize}
    \item \textit{Standard Setting}: Assuming that an exact knowledge of the measurement matrix $\boldsymbol{\Phi}$ is available at the decode, one may recover the signal of interest as follows \cite{9187438}: 
\begin{equation}
    \boldsymbol{x}_{\boldsymbol{\boldsymbol{\Phi}}}^{\star}(\boldsymbol{y}) \in \underset{\boldsymbol{x}}{\operatorname{argmin}}\ \|\rho(\boldsymbol{y} \odot(\boldsymbol{\boldsymbol{\Phi}} \boldsymbol{x}-\boldsymbol{\tau}))\|_{1}, ~~\text { s.t. }\|\boldsymbol{x}\|_{0}=k,
    \label{fun6}
\end{equation}
where $\rho(\boldsymbol{x}) = \operatorname{max}\{-\boldsymbol{x},0\}$, and is applied on the vector argument in an element-wise manner. The underlying optimization problem manifested in \eqref{fun6} is non-convex, and may be tackled by resorting to local optimization techniques. For instance, a first-order optimizer known as \textit{Binary Iterative Hard Thresholding} (BIHT)  was used in \cite{jacques2013robust}, which takes the form:
\begin{equation}
    \boldsymbol{x}_{t+1}=\mathcal{H}_k \left(\boldsymbol{x}_t+\alpha \boldsymbol{\boldsymbol{\Phi}}^T (\boldsymbol{y}-\operatorname{sign}(\boldsymbol{\boldsymbol{\Phi}} \boldsymbol{x}_t-\boldsymbol{\tau}))\right), \label{iteration}
\end{equation}
where $\alpha>0$ is the step-size and $\mathcal{H}_k(\cdot)$ is a non-linear operator that retains the largest $k$ elements (in magnitude) of the vector argument  and sets the remaining values to zero. 

\item 
\textit{Blind Setting}: 
In this scenario, which is the focus of this work, the decoder does not have any direct knowledge of the sensing matrix $\boldsymbol{\Phi}$, to directly realize \eqref{fun6}. Instead, to make the recovery possible in a data-driven manner, we assume that the decoder has access to a dataset containing the input-output pairs $\{(\boldsymbol{x}^i,\boldsymbol{y}^i)\}$, generated from the data-acquisition model.

\end{itemize}



\subsection{Problem Formulation}


Existing algorithms for CS, including the one-bit CS problem, commonly require the exact or partial\cite{fosson2020sparse} knowledge of the sensing matrix $\boldsymbol{\Phi}$ to perform the recovery task. The main idea behind blind compressive sensing, and the blind one-bit CS problem \cite{6034717}, is to circumvent requiring a prior knowledge of the measurement matrix $\boldsymbol{\Phi}$, and to indirectly facilitate learning the exact or an alternative sensing matrix from the available data at hand---so to augment the recovery algorithm. Since the noise is inevitable during signal acquisition stage, in contrast to many prior works, the measurements corrupted with a Gaussian noise are considered to secure the effectiveness of our algorithm. Such a recovery algorithm must be able to recover the signal of interest $\boldsymbol{x}$ from its one-bit noisy measurements:
  
\begin{equation}
    \boldsymbol{y}= f_\Theta(\boldsymbol{x})=\operatorname{sign}(\boldsymbol{\Phi} \boldsymbol{x} + \boldsymbol{n} - \boldsymbol{\tau}), \label{data}
\end{equation}

where $\boldsymbol{n} \sim \mathcal{N}(\mathbf{0}, \boldsymbol{C})$ denotes the additive Gaussian noise with an arbitrary covariance matrix $\boldsymbol{C}$.

All the existing recovery models are based on the assumption that the sensing matrix $\boldsymbol{\Phi}$ is known \textit{a priori} for signal reconstruction purposes or a separate signal processing and estimation stage is required to obtain an estimation of $\boldsymbol{\Phi}$. To the best of our knowledge, none of the existing works consider learning the sensing matrix as part of the decoder algorithm, i.e., learning a sensing matrix that result in the best performance of the decoder module. As such, our goal in a blind one-bit CS recovery is to reconstruct the signal of interest $\boldsymbol{x}$ from its noisy one-bit measurements $\boldsymbol{y}$ without knowledge of the underlying sparse basis $\boldsymbol{\Phi}$. In doing so, we make the following assumptions:

\begin{itemize}
    \item \textit{Data-acquisition model}, i.e., the decoder is \textit{aware} that the one-bit measurements $\boldsymbol{y}$ are generated according to the sensing model:
    \begin{equation}
        \boldsymbol{y}=\operatorname{sign}(\boldsymbol{\Phi} \boldsymbol{x} + \boldsymbol{n} - \boldsymbol{\tau})
    \end{equation}
    where the threshold vector $\boldsymbol{\tau}$ is also known at the receiver.\\
    
    \item \textit{An ensemble of input-output signal dataset} $\{\boldsymbol{x}^i,\boldsymbol{y}^i\}_{i=1}^B$ is accessible to the decoder, where B denotes the size of dataset, and we have that
    \begin{equation}
        \boldsymbol{y}^i=\operatorname{sign}(\boldsymbol{\Phi} \boldsymbol{x}^i + \boldsymbol{n} - \boldsymbol{\tau}),\ \ i\in \{1,2,...,B\}.
        \label{ensemble}
    \end{equation}
\end{itemize}

We will utilize the above assumptions to tackle the problem of blind one-bit CS in the following.


\section{Deep Unfolding Network for One-bit Compressive Sensing}
\label{DLCS}


In this section, we present our model-based deep architecture that is able to jointly \textit{optimize} the measurement sensing matrix $\boldsymbol{\Phi}$ and find an optimal decoder of the signals in order to tackle the blind one-bit CS problem. This is done as a two-stage alternating process in every deep training epoch. The first stage is devoted to finding a measurement matrix that maximizes the accuracy of the recovery algorithm through a judicious utilization of the system model and the available data at hand. The second stage is concerned with finding an optimal decoder of the signals given the obtained measurement matrix from the previous stage. All in all, the proposed methodology can be viewed as a single signal decoding framework that can learn to perform the task of recovering the underlying signal from one-bit measurements without the explicit knowledge of the measurement matrix $\boldsymbol{\Phi}$. The specific details of these two stages are described below.

\textit{\textbf{First Stage: System Identification Problem.}} 
We formally define this stage as the optimization process associated with
\begin{equation}
    \boldsymbol{\Phi}^\star = \underset{\boldsymbol{\Phi}}{\operatorname{argmin}}\sum_{i}\|\boldsymbol{x}_{\boldsymbol{\Phi}}^{\star}(\boldsymbol{y}^{i})-\boldsymbol{x}^{i}\|_2^2 \label{fun11}
\end{equation}
where $\boldsymbol{x}_{\boldsymbol{\Phi}}^{\star}$ represents the estimated signal from the reconstruction algorithm. In other words, we find a measurement matrix $\boldsymbol{\Phi}^\star$ such that it minimizes the reconstruction error between the true signal and the one recovered from the unfolded reconstruction algorithm.
Nonetheless, the underlying optimization problem manifested above is very difficult to solve. This is due to the fact that any attempt to perform the optimization with respect to the measurement matrix $\boldsymbol{\Phi}$ requires that the function $\boldsymbol{x}_{\boldsymbol{\Phi}}^{\star}$ be continuous and differentiable with respect to $\boldsymbol{\Phi}$, which is clearly not the case. In order to overcome this obstacle, we first obtain an approximation of $\boldsymbol{x}_{\boldsymbol{\Phi}}^{\star}$ by using the iterations of the algorithm presented in \eqref{iteration} such that the resulting approximation is differentiable with respect to its parameters, including $\boldsymbol{\Phi}$. More precisely, our approximated function takes the form

\begin{equation}
    g_{{\phi}_{i}}(\boldsymbol{x} ; \boldsymbol{y}, \boldsymbol{\tau})=\mathcal{H}_{k}(\boldsymbol{x}+\alpha_{i} \boldsymbol{\Phi}^{T}(\boldsymbol{y}- \operatorname{sign}(\boldsymbol{\Phi} \boldsymbol{x}-\boldsymbol{\tau}))), \label{fun12}
\end{equation}
where $\phi_i = \{\boldsymbol{\Phi}, \alpha_i\}$ denotes the set of tunable parameters. Observe that the above mapping can be viewed as a layer of a deep neural network whose activation function is given by $\mathcal{H}_k$ and its input is given by $\boldsymbol{x}$, while $(\boldsymbol{y}, \boldsymbol{\tau})$ represents the known system information. In light of this, we define $\mathcal{G}_{\Delta}^L$ as

\begin{equation}
    \mathcal{G}_{\Delta}^{L}(\boldsymbol{x}_{0} ; \boldsymbol{y}, \boldsymbol{\tau}) \triangleq g_{\boldsymbol{\Phi}_{L}} \circ g_{\boldsymbol{\Phi}_{L-1}} \circ \cdots \circ g_{\boldsymbol{\Phi}_{1}}(\boldsymbol{x}_{0} ; \boldsymbol{y}, \boldsymbol{\tau}),
    \label{x1}
\end{equation}
where $\Delta=\bigcup_{i} \phi_i$ denotes the global parameters of the system. In the following, we delve deeper by discussing two distinct, although connected perspectives on \eqref{x1}.

$\bullet$ \textit{\textbf{Optimization Theory Perspective:}} 
The mapping \eqref{x1} can be viewed as performing L first-order iterations of the form \eqref{fun12} which is basically designed to solve the optimization problem manifested in \eqref{fun6}. The mapping function $\mathcal{G}$ is parameterized not only on the per-iteration step-sizes $\{\alpha_i\}$, but also on the measurement matrix $\boldsymbol{\Phi}$. Moreover, note that for a fixed $\alpha_{L}=\alpha_{L-1}=\cdots=\alpha_{1}=\alpha$, and some proper choice of $\alpha$, the function $\mathcal{G}_{\Delta}^L$ approaches the first-order stationary point of the objective function with limited error. It should be clear at this point that the mapping $\mathcal{G}_{\Delta}$ provides an approximation of $\boldsymbol{x}_{\boldsymbol{\Phi}}^\star$, for a proper choice of $L$ and $\alpha$. Accordingly, we propose to regard the measurement matrix $\boldsymbol{\Phi}$ as a trainable parameter of the function $\mathcal{G}_{\Delta}^L$ and seek to learn it by exploiting the available data, i.e., solving \eqref{fun11}. On the other hand, once a solution $\boldsymbol{\Phi}^\star$ is obtained, we regard $\{\alpha_i\}$ as the set of trainable parameters of the mapping and perform another round of training to accelerate the convergence of the underlying mapping to the true signal.




\begin{algorithm}[t]
\caption{Training Procedure for Proposed Method}
\textbf{Initialize} $\boldsymbol{\Phi}(0)$ and $\alpha_i(0)$\\
    \For{epoch = 1,...,N}{
        \For{k = 1,...,L}{
            Feed $\boldsymbol{x}(k)$ to the k-th layer of the DNN \eqref{x1}\\
        }
        Compute the loss function \eqref{loss}\\
        Utilize back propagation to update parameters $\Delta$
        }
\end{algorithm}

$\bullet$ \textit{\textbf{Deep Learning Perspective:}} The mapping \eqref{x1} can be further viewed as a L-layer feed forward neural network whose input is given by an initial point $\boldsymbol{x_0}$, the one-bit measurements $\boldsymbol{y}$, and the quantization thresholds $\boldsymbol{\tau}$. Unlike the traditional black-box data-driven approaches, the resulting neural network $\mathcal{G}_{\Delta}^L$ is tailored to the problem at hand and is now interpretable. Due to the incorporation of domain-knowledge into the architecture and computational dynamics of the proposed neural network, it is expected to learn to decode a $K$-sparse signal without an explicit knowledge of $\boldsymbol{\Phi}$, with significantly smaller number of training data points and trainable parameters.

In light of the above discussion, we reform \eqref{fun11} into the following tractable form:

\begin{equation}
    \mathbf{\boldsymbol{\Phi}}^{\star}=\underset{\mathbf{\boldsymbol{\Phi}}}{\operatorname{argmin}} \sum_{i}\left\|\mathcal{G}_{\boldsymbol{\Delta}=\left\{\boldsymbol{\alpha}_{i}=\alpha, \boldsymbol{\boldsymbol{\Phi}}\right\}_{i}}^{L}\left(\boldsymbol{x}_{0} ; \boldsymbol{y}^{i}, \boldsymbol{\boldsymbol{\tau}}\right)-\boldsymbol{x}^{i}\right\|_{2}^{2}.
    \label{fun14}
\end{equation}
Hence, obtaining $\boldsymbol{\Phi}^\star$ via \eqref{fun14} can be viewed as training the proposed deep neural network $\mathcal{G}_{\Delta}^L$ over the parameter $\boldsymbol{\Phi}$.

\textit{\textbf{Second Stage: Learning to Optimize.}} In this part, we discuss the second training stage of the proposed methodology, which corresponds to learning a set of step-size scalars $\{\alpha_i\}$ such that the resulting network converges to an stationary point of the optimization problem in $L'$ iterations \cite{takabe2020theoretical}. We note that during the first training stage, we usually choose a large value for $L$ to ensure the convergence of $\mathcal{G}_\Delta$ to a stationary point, and hence, facilitating the inference of the measurement matrix. Once $\boldsymbol{\Phi}^\star$ is obtained through \eqref{fun14}, we reduce the number of layers of the neural network from $L$ to $L'$, i.e. $L'\leq L$, and perform the training of the neural network on the set of parameters $\{\alpha_i\}_{i=1}^{L'}$, while keeping $\boldsymbol{\Phi}^\star$ fixed.

Mathematically speaking, this stage corresponds to performing the training according to:
\begin{equation}
    \min _{\left\{\alpha_{i}\right\}_{i}} \sum_{i}\left\|\mathcal{G}_{\boldsymbol{\Delta}=\left\{\boldsymbol{\alpha}_{i}=\alpha, \boldsymbol{\boldsymbol{\Phi}}^{\star}\right\}_{i}}^{L'}\left(\boldsymbol{x}_{0} ; \boldsymbol{y}^{i}, \boldsymbol{\boldsymbol{\tau}}\right)-\boldsymbol{x}^{i}\right\|_{2}^{2}.
\end{equation}
In this training process design, one should consider the constraint on the step sizes set $\{\alpha_i\}_{i=1}^{L'}$, which should be always non-negative. Hence, it is necessary to apply regularization in the training loss function to facilitate choosing positive step sizes and to avoid overfitting. Accordingly, the loss function we suggest for training is as follows:
\begin{equation}
\begin{split}
    LOSS= &\underbrace{\sum_{i=1}^{L'} \left\|\hat{\mathcal{G}}_{\boldsymbol{\Delta}}^{L'}\left(\boldsymbol{x}_{0};\boldsymbol{y}^i,\boldsymbol{\tau}\right)-\boldsymbol{x}^{i}\right\|_{2}^{2}}_{\text {accumulated MSE loss of all layers }}\\
    + &\underbrace{\lambda \sum_{i=1}^{L'} \operatorname{ReLU} \left(-\alpha_{i}\right)}_{\text {regularization term for the step sizes}}
\end{split} \label{loss}
\end{equation}
where ReLU is the Rectified Linear Unit activation function, which can also help to avoid overfitting problem as a regularizer \cite{glorot2011deep}.

Once both training stages are carried out successfully, the proposed neural network $\mathcal{G}_{\Delta}^{L^{\prime}}$ can be used to decode a $K$-sparse signal from its one-bit measurements.


\begin{figure}[t]
\centerline{\includegraphics[width=0.8\columnwidth]{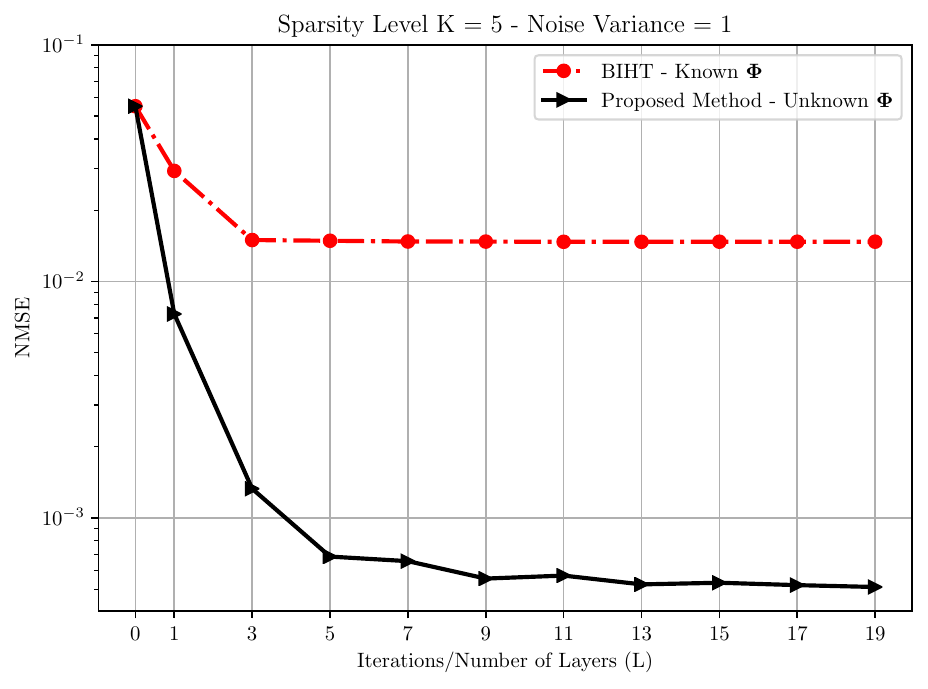}}
\caption{The performance of the proposed model compared to the BIHT Method for growing iterations/number of layers. }
\end{figure}

\begin{figure}[t]
\centerline{\includegraphics[width=0.8\columnwidth]{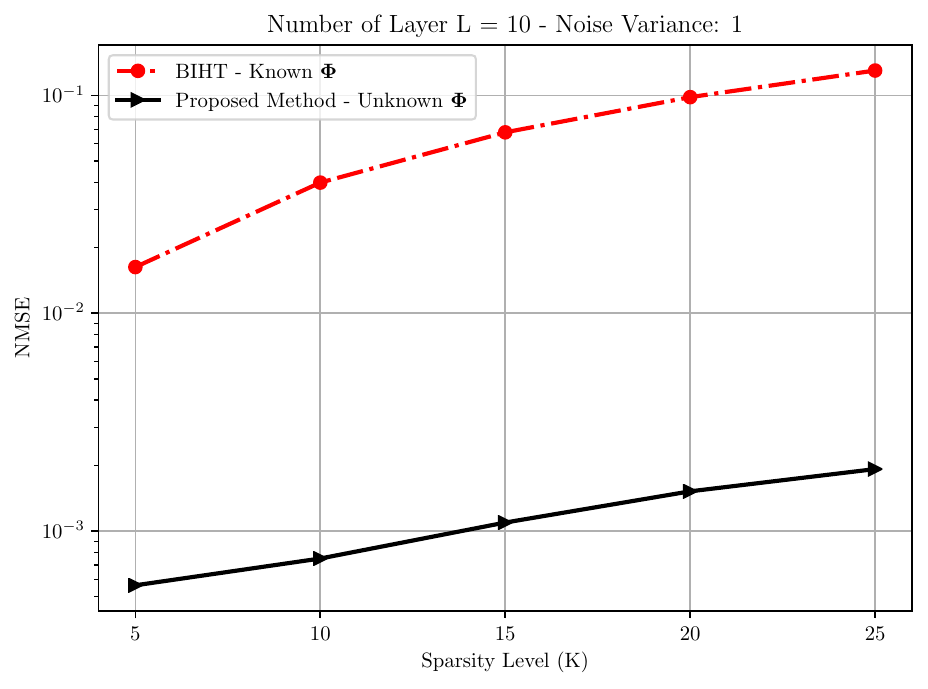}}
\caption{The performance of the proposed model compared to the BIHT Method for a growing number of sparsity levels}
\end{figure}


\section{Numerical results}
In this section, we present our simulation results for the proposed blind one-bit CS methodology in the presence of the measurement noise in order to show the effectiveness of the training in our deep model and its inference. The simulations are implemented and performed using the PyTorch library \cite{paszke2019pytorch} and the parameter optimization process is done on the Adam optimizer\cite{8624183} with a learning rate of $10^{-4}$. For training, the non-zero elements of K-sparse signals of length $n=128$ are randomly generated from $\mathcal{N}(0, 1)$ using Xavier initialization \cite{kumar2017weight}, and the additive Gaussian noise is generated with unit variance. Furthermore, the number of layers of the proposed network model is fixed at $L=10$. For the sensing matrix training, we assume $\boldsymbol{\Phi} \in \mathcal{R}^{512\times128}$, and all initial elements in the matrix are generated from a standard normal distribution. We develop the framework to handle arbitrary user-specified quantization thresholds. For the numerical experiments, we consider the more commonly-used case of $\boldsymbol{\tau} = \mathbf{0}$. Since the gradient of sign function in \eqref{fun12} is zero almost everywhere, straight-through estimator (STE) \cite{QIN2020107281} is implemented in the proposed model to back propagate through the binary function sign. Also, a normalization step that corresponds to implementing a projected gradient descent step is implemented at the end of each layer in order to help the algorithm converge. These two results presented are averaged over 20 realizations of the system parameters.


Figure 1 illustrates NMSE between the recovered signal and the true underlying signal, for the output of each layer, for the proposed methodology and the baseline BIHT method. The initial starting point of these two methods are the same. It can be observed that the proposed deep model is indeed reducing the NMSE per layer/iteration and the proposed methodology results in accelerated convergence properties. 
We note that such per-layer analysis is only possible due to the model-based nature of the proposed deep architecture. 

Figure 2 illustrates the performance of the proposed model and the original BIHT method in terms of NMSE for various sparsity levels (the length of the $K$-sparse signal is fixed at $n = 128$). It can be observed from Fig. 2 that the proposed methodology exhibits a superior performance when compared to the state-of-the-art BIHT method yielding a slower degradation in performance as compared to the baseline BIHT algorithm, as sparsity level $K$ grows larger. 

It is worth mentioning that the proposed methodology is performing the task of recovery in a blind manner (no access to the underlying sensing matrix), while the BIHT methodology exploits a full knowledge of $\boldsymbol{\Phi}$. The superior performance of the proposed methodology is related to the fact that in noisy scenarios, learning a surrogate sensing matrix can indeed be beneficial and can result in more robust iterations as compared to the base-line BIHT method. 
In addition, learning the step-size scalars for each layer is helpful for the proposed architecture to escape many poor local minima.

\section{CONCLUSION}
In this work, we considered the problem of blind one-bit CS recovery. In particular, we assumed that the reconstruction algorithm does not have the knowledge of the underlying sensing matrix, and proposed a novel hybrid data-driven and model-based technique for the problem at hand. Our simulations demonstrate that our model-based deep neural networks go beyond the performance of the conventional base-line algorithm while requiring a relatively low number of training parameters---which means both a high accuracy of recovery and efficiency in data utilization.

\bibliographystyle{ieeetr} 
\bibliography{ref}

\end{document}